\documentclass[%
 reprint,
 amsmath,amssymb,
 aps
pra,prb
]{revtex4}

\usepackage{graphicx}
\usepackage{dcolumn}
\usepackage{bm}
\usepackage{hyperref}
\usepackage[mathlines]{lineno}

\newcommand{\R}{{\mathbb R}}
\newcommand{\dem}{{\em Proof: \;}}
\newcommand{\fdem}{\hfill $\square$}

\def\noi{\noindent}

\newtheorem{teo}{Theorem}[section]

\newtheorem{prop}[teo]{Proposition}

\newtheorem{defi}{Definition}[section]

\begin{document}


\title{Do M\"{a}rzke-Wheeler effects influence on measured data in nature?}

\author{Juan Manuel Burgos}

\altaffiliation{Instituto de Matem\'aticas, Universidad Nacional Aut\'onoma de M\'exico, Unidad Cuernavaca.\\ Av. Universidad s/n, Col. Lomas de
Chamilpa. Cuernavaca, Morelos M\'exico, 62209.}
\email{burgos@matcuer.unam.mx}

\date{\today}

\begin{abstract}
\noi We wonder whether M\"{a}rzke-Wheeler effects influence on measured data in nature. Through a formula developed in this letter for the calculation of the M\"{a}rzke-Wheeler map of a general accelerated observer, we study the influence of the M\"{a}rzke-Wheeler acceleration effect on the NASA's Pioneer anomaly and found that it is about a fifth of the anomaly value. Due to statistical errors in the measured anomaly, it is not possible to neither confirm nor neglect the influence of the M\"{a}rzke-Wheeler acceleration effect on the measured Pioneer data. We hope that the ideas presented here could encourage other research teams in the search for other observational objects that could finally answer the question posed in this letter.
\end{abstract}


\maketitle

\section{Introduction}

\noi Nowadays, it is very clear how special relativity effects influence on measured data. The first celebrated example of this fact was the atmospheric muons decay explanation as a time dilation effect. This is the Rossi-Hall experiment \cite{RH}. Considering the M\"{a}rzke-Wheeler synchronization \cite{MW} as the natural generalization to accelerated observers of Einstein synchronization in special relativity, we wonder whether M\"{a}rzke-Wheeler effects influence on measured data in nature. This question is also motivated by the fact that recently the twin paradox was completely solved in (1+1)-spacetime by means of these effects \cite{Bu} and it is natural to ask for empirical confirmation. Of course these effects comprehend the well known special relativistic ones for inertial observers as well as the new ones. These new effects can be seen as corrections of the special relativistic ones due to the acceleration of the involved observer.
\\

\noi A small deviation towards the sun from the predicted Pioneer acceleration: $8.09 \pm 0.20 \times 10^{-10} m/s^{2}$ for Pioneer 10 and $8.56 \pm 0.15 \times 10^{-10} m/s^{2}$ for Pioneer 11, was reported for the first time in \cite{An98}. The analysis of the Pioneer data from 1987 to 1998 for Pioneer 10 and 1987 to 1990 for Pioneer 11 made in \cite{An05} improves the anomaly value and it was reported to be $8.74 \pm 1.33\times 10^{-10} m/s^{2}$. This is known as the Pioneer anomaly.
\\

\noi Considering that M\"{a}rzke-Wheeler tiny effects are difficult to measure, we careful looked for some observational object for which the searched effect could be appreciable. This search led us to the Pioneer 10. In fact, through a simple analytic formula for the M\"{a}rzke-Wheeler map exact calculation developed in this letter, computing the acceleration difference between the M\"{a}rzke-Wheeler and Frenet-Serret coordinates for the earth's translation around the sun, we see that this M\"{a}rzke-Wheeler long range effect is between $0$ and $\approx 17\%$ of the Pioneer anomaly value. Unfortunately, due to statistical errors in the measured anomaly, it is not possible to confirm the influence of the M\"{a}rzke-Wheeler acceleration effect on the measured Pioneer data. Moreover, a recently numerical thermal model based on a finite element method \cite{Tu12} has shown a discrepancy of $20\%$ of the actual measured anomaly and due to the mentioned statistical errors, it was concluded there that the pioneer anomaly has been finally explained within experimental error of $26\%$ of the anomaly value:
\\

\emph{
...To determine if the remaining $20\%$ represents a statistically significant acceleration anomaly not accounted for by conventional forces, we analyzed the various error sources that contribute to the uncertainties in the acceleration estimates using radio-metric Doppler and thermal models... We therefore conclude that at the present level of our knowledge of the Pioneer 10 spacecraft and its trajectory, no statistically significant acceleration anomaly exists.}
\\

\noi Although it is tempting to think that the $20\%$ discrepancy found in \cite{Tu12} is due to a long range M\"{a}rzke-Wheeler acceleration effect, it cannot be confirmed. We hope that the ideas presented here could encourage other research teams in the search for other observational objects that could finally answer the question posed in this letter.
\\

\section{M\"{a}rzke-Wheeler map}

\noi Consider the $(1+n)$-spacetime $\mathcal{M}$ spanned by the vectors $\sigma_{0},\sigma_{1}\ldots \sigma_{n}$ with the Lorentz metric: $$ds^{2}= (dx^{0})^{2}-(dx^{1})^{2}-\ldots (dx^{n})^{2}$$ respect to the basis $\{\sigma_{0},\sigma_{1}\ldots \sigma_{n}\}$. An observer is a smooth curve $\gamma: \R\rightarrow \mathcal{M}$ naturally parameterized with timelike derivative at every instant; i.e. $|\dot{\gamma}(s)|^{2}=1$. We will say a vector is spatial if it is a linear combination of $\{\sigma_{1}\ldots \sigma_{n}\}$. A spatial vector $\vec{u}$ is unitary if $|\vec{u}|^{2}=-1$.
\\
\begin{defi}
\noi Consider a timelike vector $a$ in $\mathcal{M}$; i.e. $|a|_{L}^{2}\geq 0$. We define the scaled Lorentz transformation $\mathcal{L}(a)$: $$\mathcal{L}(a)= |a|_{L} \mathcal{L}_{a}$$ where $\mathcal{L}_{a}$ is the orthocronous Lorentz boost transformation sending $\sigma_{0}$ to the unitary vector $a/|a|_{L}$; i.e. the original and transformed coordinates are in standard configuration ($x'$, $y'$ and $z'$ are colinear with $x$, $y$ and $z$ respectively where the prime denote the spatial transformed coordinates and the others denote the original spatial coordinates).
\end{defi}

\begin{figure}
\begin{center}
  \includegraphics[height=0.2\textwidth]{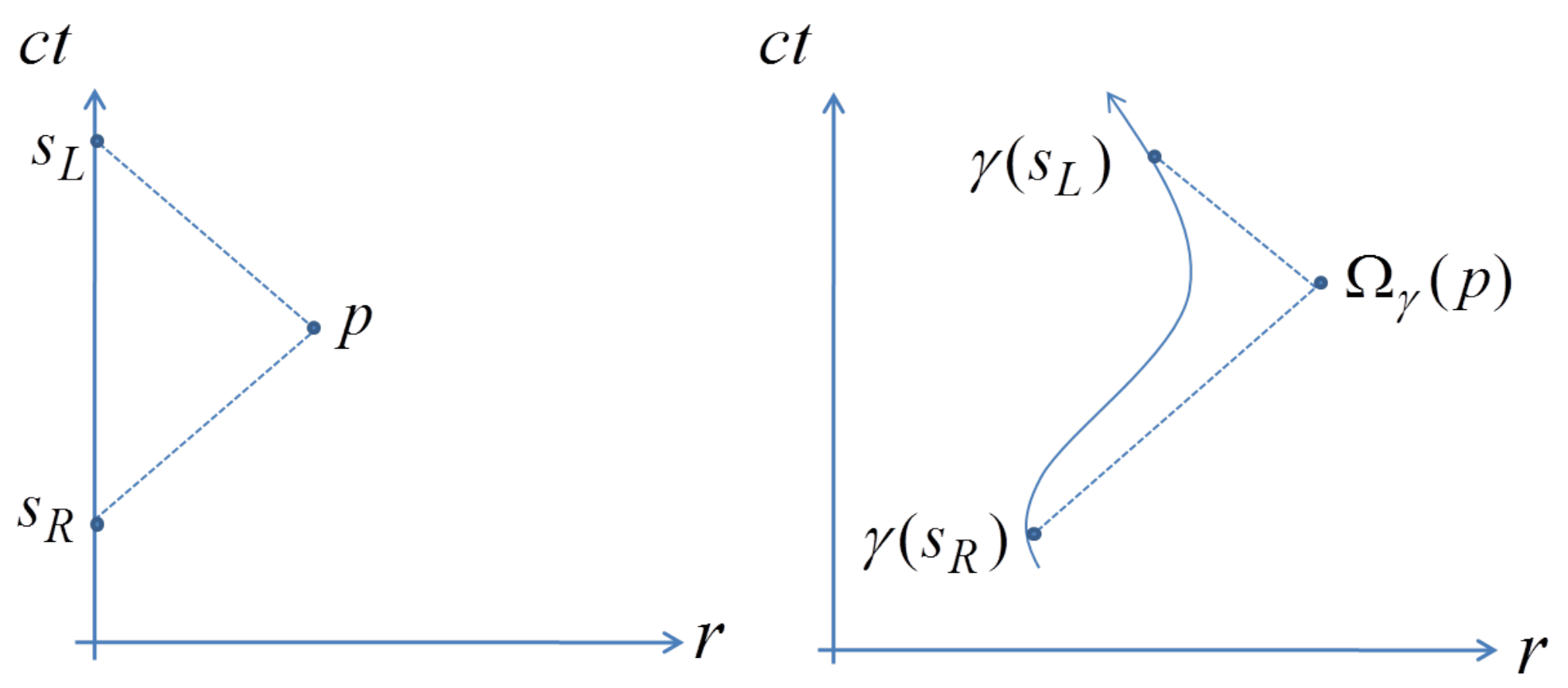}\\
  \end{center}
  \caption{M\"{a}rzke-Wheeler map}\label{MW_Coord}
\end{figure}

\noi The scaled Lorentz transformation has the following properties: $$|\mathcal{L}(a)(b)|_{L}^{2}=|a|_{L}^{2}\ |b|_{L}^{2}$$ $$\mathcal{L}(a)(\sigma_{0})= a$$
\\
\begin{defi}
\noi A smooth map $\Omega_{\gamma}:\mathcal{M}\rightarrow \mathcal{M}$ is a M\"{a}rzke-Wheeler map of the observer $\gamma$ if it verifies: $$|\Omega_{\gamma}(s\ \sigma_{0}+r \vec{u})-\gamma(s\pm r)|_{L}^{2}=0$$
for every real $s$, positive real $r$ and unitary spatial vector $\vec{u}$ (see Figure \ref{MW_Coord}).
\end{defi}

\noi This map \cite{MW}, \cite{PV}, \cite{Bu} is clearly an extension of the Einstein synchronization convention for non accelerated observers; i.e. It is the natural generalization of a Lorentz transformation in the case of accelerated observers.

\begin{prop}\label{MWformula}
\noi Consider an observer $\gamma: \R\rightarrow \mathcal{M}$. Then,
\begin{eqnarray*}
  \Omega_{\gamma}(s\ \sigma_{0}+r\vec{u}) &=& \frac{\gamma(s+r)+\gamma(s-r)}{2} \\
   & & +\ \mathcal{L}\left(\frac{\gamma(s+r)-\gamma(s-r)}{2}\right)(\vec{u})
\end{eqnarray*}
is a M\"{a}rzke-Wheeler map of the observer $\gamma$ such that $\vec{u}$ is a unitary spatial vector.
\end{prop}
\dem
\noi Recall that for every $a$ such that $|a|_{L}^{2}\geq 0$ we have that $\mathcal{L}(a)(\sigma_{0})=a$. This way,

\begin{eqnarray*}
& &  |\Omega_{\gamma}(s\ \sigma_{0}+r\vec{u})-\gamma(s\pm r)|_{L}^{2}\ =\  |\mp\frac{\gamma(s+r)-\gamma(s-r)}{2} \\
& & +\ \mathcal{L}\left( \frac{\gamma(s+r)-\gamma(s-r)}{2}  \right)(\vec{u})|_{L}^{2} \\
&=& |\mathcal{L}\left(\frac{\gamma(s+r)-\gamma(s-r)}{2}\right)(\vec{u}\mp \sigma_{0})|_{L}^{2} \\
&=& |\frac{\gamma(s+r)-\gamma(s-r)}{2}|_{L}^{2}\ |\vec{u}\mp \sigma_{0}|_{L}^{2}=0
\end{eqnarray*}
because $|(\vec{u}\mp \sigma_{0})|_{L}^{2}=0$. From the formula it is clear that $\Omega_{\gamma}$ is smooth.
\fdem
\\

\noi The last M\"{a}rzke-Wheeler map formula was written for the first time in \cite{Bu} for $(1+1)$-spacetime where it was shown, in this particular case, that it is actually a conformal map. Moreover, the twin paradox is solved in $(1+1)$-spacetime. In the general case treated here, the M\"{a}rzke-Wheeler map is no longer conformal.
\\

\noi As an example, consider the uniformly accelerated observer in $(1+3)$-spacetime along the $\sigma_{1}$ axis: $$\gamma(s)= p+ R\left(\sinh \left(\frac{s}{R}\right)\sigma_{0} + \cosh \left(\frac{s}{R}\right)\sigma_{1}\right)$$
where $s$ is its natural parameter and $R=c^{2}/a$ such that $a$ is the observer acceleration. Its M\"{a}rzke-Wheeler map is:

\begin{eqnarray*}
\Omega_{\gamma}(s, x, y, z)  &=& p+ R\sinh \left(\frac{s}{R}\right)\left[\cosh \left(\frac{s}{R}\right) + \sinh \left(\frac{s}{R}\right)\frac{x}{r}\right]\sigma_{0} \\
& &       + R\cosh \left(\frac{s}{R}\right)\left[\cosh \left(\frac{s}{R}\right) + \sinh \left(\frac{s}{R}\right)\frac{x}{r}\right]\sigma_{1} \\
& &      + \frac{R}{r}\sinh \left(\frac{r}{R}\right)\left[ y\ \sigma_{2} + z\ \sigma_{3} \right] \\
\end{eqnarray*}
where $r^{2}= x^{2} + y^{2} + z^{2}$. In this example, it is interesting that besides $\Omega_{\gamma}$ restricted to the $\langle\ \sigma_{0},\ \sigma_{1}\ \rangle$ plane is a conformal map (as it was expected from \cite{Bu}), it is also also conformal restricted to the $\langle\ \sigma_{2},\ \sigma_{3}\ \rangle$ plane.

\section{Pioneer anomaly}

\noi The Pioneer 10/11 data is measured from Earth's DSN antennas (Deep Space Network) and we wonder whether this data is affected by earth's translation around the sun. We comment about earth's rotation at the end of the section. We model the Earth's translation as the uniformly rotating observer $$\gamma(s)= \frac{s}{k}\sigma_{0}+ R\left[ \cos\left(\frac{\omega}{ck}s\right)\sigma_{1} + \sin\left(\frac{\omega}{ck}s\right)\sigma_{2}\right]$$ where $s$ is its natural parameter, $k=\sqrt{1-R^{2}\omega^{2}/c^{2}}$ is its Lorentz contraction factor and $R|\omega|<c$. Its M\"{a}rzke-Wheeler map is:

\begin{eqnarray*}
\Omega_{\gamma}(s, x, y, z)  &=& \left[\frac{s}{k} +  \frac{R}{r}\sin\left(\frac{\omega}{ck}r\right)y\right]\sigma_{0} \\
& &       + \left[ R\cos\left(\frac{\omega}{ck}r\right) + x\ \sqrt{\frac{1}{k^{2}} - \left(\frac{R}{r}\sin\left(\frac{\omega}{ck}r\right)\right)^{2}} \right]\vec{a}(s) \\
& &     + \frac{1}{k}\ y\ \vec{b}(s) \\
& &		+ z\ \sqrt{\frac{1}{k^{2}} - \left(\frac{R}{r}\sin\left(\frac{\omega}{ck}r\right)\right)^{2}}\ \sigma_{3} \\
\end{eqnarray*}
where $r^{2}= x^{2} + y^{2} + z^{2}$. We have chosen the framing $\{\vec{a}, \vec{b}, \sigma_{3}\}$ corresponding to the $x,y,z$ coordinates such that $\{\ \vec{b},\ -\vec{a},\ \sigma_{3}\ \}$ is the Frenet-Serret framing of the observer (see Figure \ref{FrenetSerret}):
$$a(s)= \cos\left(\frac{\omega}{ck}s\right)\sigma_{1} + \sin\left(\frac{\omega}{ck}s\right)\sigma_{2}$$
$$b(s)=  -\sin\left(\frac{\omega}{ck}s\right)\sigma_{1} + \cos\left(\frac{\omega}{ck}s\right)\sigma_{2}$$
This expression was also obtained in \cite{PV} in the particular case $z=0$. It is interesting to notice the oscillatory term of the above map. In order to compare the spatial M\"{a}rzke-Wheeler coordinates with the Frenet-Serret coordinates we consider the difference $\Omega_{\gamma}(s, x, y, z) - \gamma(s)$. Because $\omega=2 \pi /year$ and $R=1 AU$, we have that $\omega R/c \approx 10^{-4}$ and restricted to the region $r<< c/\omega \approx 10^{4}AU$ we have the approximation:

\begin{figure}
\begin{center}
  \includegraphics[height=0.2\textwidth]{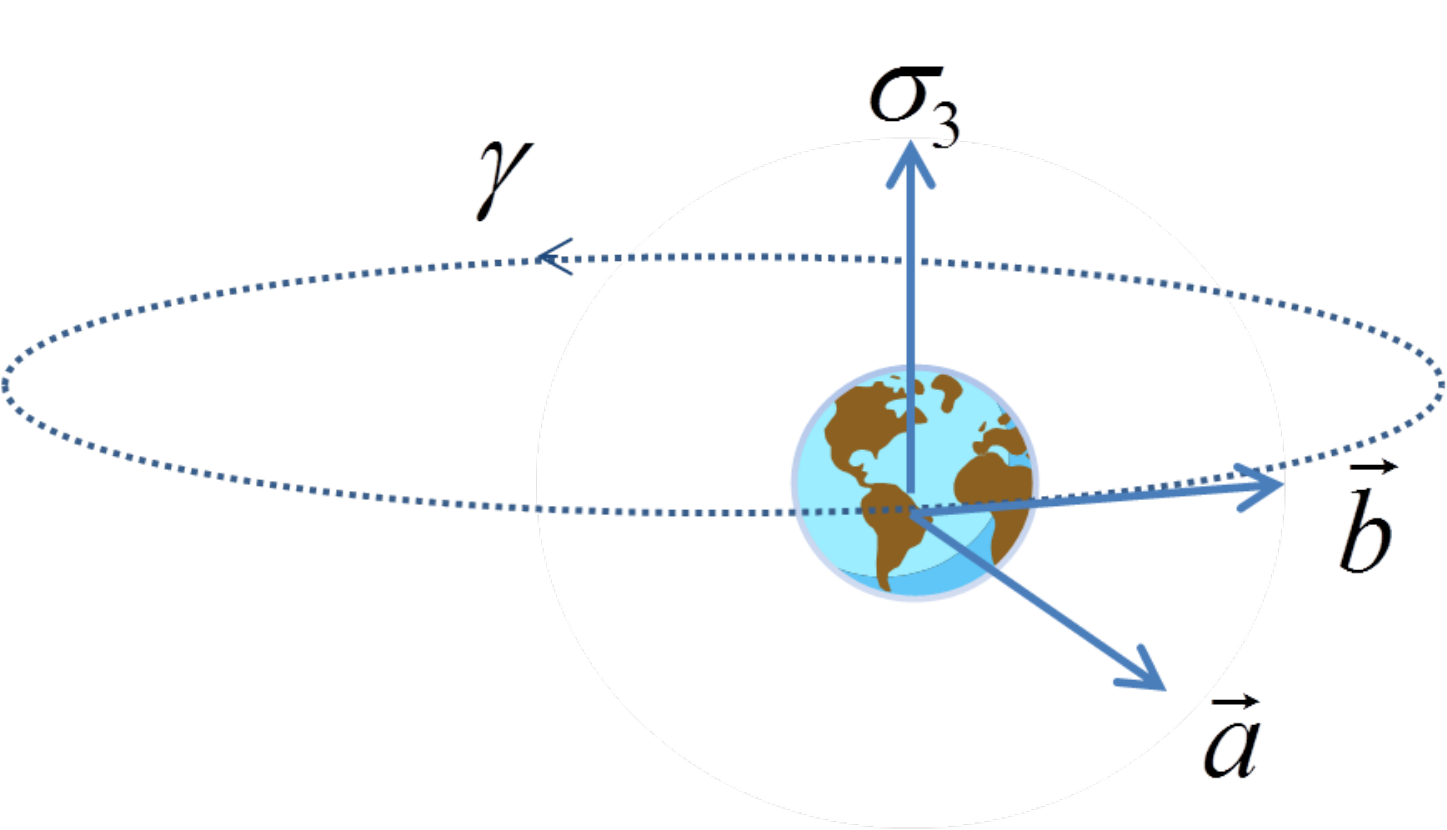}\\
  \end{center}
  \caption{Observer's framing}\label{FrenetSerret}
\end{figure}

\begin{eqnarray*}
\Omega_{\gamma}(s, x, y, z)   - \gamma(s) &=&  \left[ R\left(\cos\left(\frac{\omega}{c}r\right)-1\right) + x\right]\ \vec{a}(s) \\
& &	+\ y\ \vec{b}(s)+ z\ \sigma_{3} \\
\end{eqnarray*}
Because the $\sigma_{0}$ component is zero, we have the following transformation between the spatial M\"{a}rzke-Wheeler coordinates and the Frenet-Serret coordinates:

\begin{eqnarray*}
x' &=&  R\left(\cos\left(\frac{\omega}{c}r\right)-1\right) + x \\
y' &=&	y \\
z' &=&	z \\
\end{eqnarray*}
where $r^{2}= x^{2} +y^{2} +z^{2}$. Because the Pioneer's velocity and acceleration are very small respect to the natural scale of the problem $c/\omega\approx 10^{4}AU$, differentiating the above expression we have:

\begin{eqnarray*}
a_{x}' &=&  a_{x} - R\frac{\omega^{2}}{c^{2}}\cos\left(\frac{\omega}{c}r\right)(\dot{r})^{2} - R\frac{\omega}{c}\sin\left(\frac{\omega}{c}r\right)\ddot{r} \\
a_{y}' &=&	a_{y} \\
a_{z}' &=&	a_{z} \\
\end{eqnarray*}where $r$ is the distance from the sun and $a$ is the acceleration.
Because the recorded Pioneer data (at least for Pioneer 10) corresponds to the region between $1AU$ and $\approx 80AU$, we can consider that $r_{Pioneer}<<c/\omega \approx 10^{4}AU$ where $r_{Pioneer}$ is the Pioneer's distance from the sun and we have the following approximation:

\begin{eqnarray*}
a_{x}' &=&  a_{x} - R\frac{\omega^{2}}{c^{2}}\ v^{2}\ \cos^{2}\varphi \\
a_{y}' &=&	a_{y} \\
a_{z}' &=&	a_{z} \\
\end{eqnarray*}
where $v$ is the Pioneer's speed and $\varphi$ is the angle between its radius vector from the sun and its velocity vector. Computing the acceleration difference $\Delta a_{x}$ between the M\"{a}rzke-Wheeler and Frenet-Serret coordinates at the Pioneer's maximal speed $v_{max}=48.000\ m/s$, we have the result:
$$0\leq |\Delta a_{x}| \leq R\ \frac{\omega^{2}}{c^{2}}\ v^{2}\approx 1.5\times 10^{-10} m/s^{2}$$
and we see that it is between $0$ and $\approx 17 \%$ of the measured Pioneer anomaly $a_{p}=8.74 \pm 1.33\times 10^{-10} m/s^{2}$.
\\

\noi The calculated difference $\Delta a_{x}$ points towards the $z$ edge when $x>0$ and in the opposite direction when $x<0$. This would contradict the claim that the anomaly always points towards the sun made in the data analysis \cite{An98} and \cite{An05}. However, in the data analysis made in \cite{Tu11}, it is claimed that it cannot be confirmed whether the anomaly is sunwards, contrary to the earlier claim.
\\

\noi Finally, we would like to comment about a possible numerical analysis on the influence of Earth's rotation on the measured data. In order to do so, we define the following framing dependent non abelian product of observers: $$\gamma \cdot \xi= \Omega_{\gamma}\circ \xi$$ This product is the generalization of the special relativistic velocities addition and has the following property:
$$\Omega_{\gamma \cdot \xi}= \Omega_{\gamma }\circ \Omega_{\xi}$$
This way, the observer $\gamma= \gamma_{Translation}\cdot\gamma_{Rotation}$ is the one we should consider and its M\"{a}rzke-Wheeler map is just the composition of the previously exactly calculated map of the uniformly rotating observer. Unfortunately, the map gets really involved and the analysis must be done numerically. An analysis of the parameters involved in the rotation analysis, shows that the magnitude order of the long range M\"{a}rzke-Wheeler acceleration effect coincides with the one of the Pioneer anomaly and should also be considered.

\section{Conclusion}

\noi Although after strongly numerical evidence it is tempting to think that the $20\%$ discrepancy of the anomaly value found in \cite{Tu12} is due to a long range M\"{a}rzke-Wheeler acceleration effect described in this letter, due to statistical errors in the measured anomaly it cannot be neither confirmed nor neglected. We hope that the ideas presented here could encourage other research teams in the search for other observational objects that could finally answer whether M\"{a}rzke-Wheeler effects influence on measured data in nature.



\begin{thebibliography}{99}

\bibitem{An98}
Anderson et al., {\it Indication, from Pioneer 10/11, Galileo, and Ulysses Data, of an Apparent Anomalous, Weak, Long-Range Acceleration\/}, Phys.Rev.Lett. 81 (1998) 2858-2861

\bibitem{An05}
Anderson et al., {\it Study of the anomalous acceleration of Pioneer 10 and 11\/} Phys.Rev.D 65: 082004, 2002.

\bibitem{Bu}
Burgos J, {\it Two-dimensional Minkowski causal automorphisms and conformal maps \/} Class. Quantum Grav. 30 (2013) 035007.

\bibitem{MW}
M\"{a}rzke M, Wheeler J, {\it Gravitation as geometry - I: the geometry of Spacetime and the geometrodynamical standard meter\/} Gravitation and relativity, H.-Y. Chiu and W. F. Hoffmann, eds.,  W. A. Benjamin, New York-Amsterdam (1964) 40.

\bibitem{PV}
Pauri M, Vallisneri M, {\it M\"{a}rzke-Wheeler coordinates for accelerated observers in special relativity\/} Found. Phys. Lett. \textbf{13} (2000) 401.

\bibitem{RH}
Rossi B, Hall D.B, {\it Variation of the Rate of Decay of Mesotrons with Momentum\/} Phys.Rev. 59, 101103 (1941).

\bibitem{Tu12}
Turyshev et al., {\it Support for the thermal origin of the Pioneer anomaly \/}  Phys. Rev. Lett. 108, 241101 (2012)

\bibitem{Tu11}
Turyshev et al., {\it Support for temporally varying behavior of the Pioneer anomaly from the extended Pioneer 10 and 11 Doppler data sets\/} Phys. Rev. Lett. 107, 081103 (2011).



\end{thebibliography}
\end{document}